\documentclass[%
reprint,
nofootinbib,
 amsmath,amssymb,
 aps,
]{revtex4-2}

\usepackage{graphicx}
\usepackage{dcolumn}
\usepackage{bm}

\begin{document}

\preprint{APS/123-QED}

\title{Wide-bandgap optical materials for high-harmonics generation at
  the nanoscale}


\author{Albert Mathew$^1$, Sergey Kruk$^{*1}$, Shunsuke Yamada$^2$, Kazuhiro Yabana$^3$, Anatoli Kheifets$^1$}

\affiliation{%
$^1$Research School of Physics, The Australian National University, Canberra, ACT, 2600, Australia}%

\affiliation{%
$^2$Kansai Photon Science Institute, National Institutes for Quantum Science and Technology, Kyoto 619-0215, Japan}%

\affiliation{%
$^3$Center for Computational Sciences, University of Tsukuba, Tsukuba 305-8577, Japan}%
\email{sergey.kruk@anu.edu.au}
\date{\today}

\begin{abstract}
High-order harmonics generation (HHG) is the only process that enables
table-top-size sources of extreme-ultraviolet (XUV) light. The HHG
process typically involves light interactions with gases or plasma --
material phases that hinder wider adoption of such sources. This
motivates the research in HHG from nanostructured solids. Here we
investigate theoretically material platforms for HHG at the nanoscale
using first-principle supercomputer simulations. We reveal that
wide-bandgap semiconductors, aluminium nitride AlN and silicon nitride
SiN, are highly-promising for XUV light generation when compared to
one of the most common nonlinear nanophotonic material -- silicon. In
our calculations we assume excitation with 100 fs pulse duration,
1$\times$10$^{13}$W/cm$^2$ peak power and 800 nm central wavelenght. We demonstrate
that in AlN material the interplay between the crystal symmetry and
the incident light direction and polarization can enable the
generation of both even and odd harmonics. Our results should advance
the developments of high-harmonics generation of XUV
light from nanostructured solids.
\end{abstract}

\maketitle
\section{Introduction}

\begin{figure}[ht]
\centering
\includegraphics[width=\linewidth]{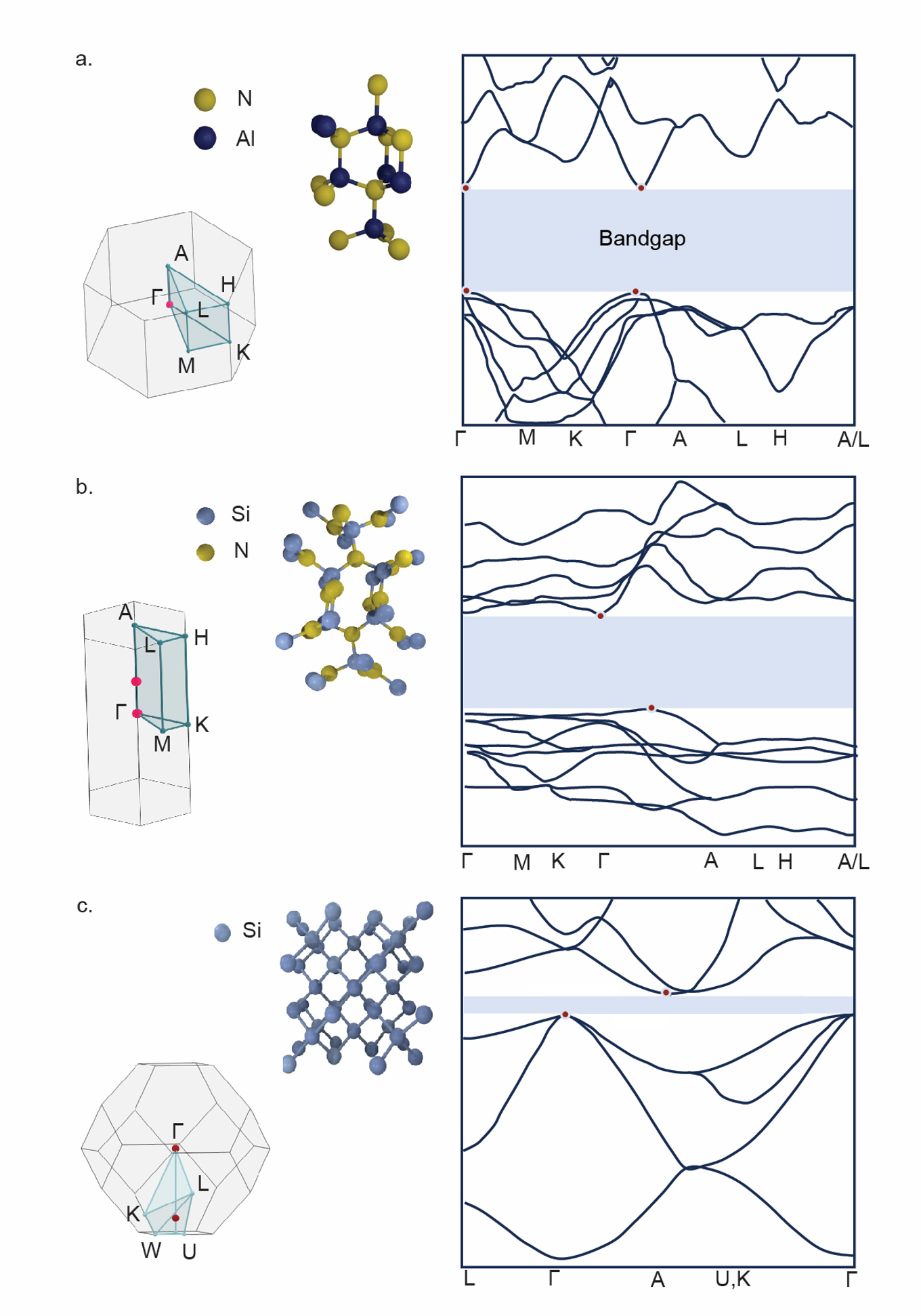}
\caption{Left: the crystalline structure of $w$-AlN (a), $\beta$-SiN (b) and Si (c) is illustrated in the coordinate and reciprocal spaces.  Right: the electronic structure of these materials is depicted in several high-symmetry directions. The band gap separating the top of the valence band and the bottom of the conduction band is highlighted.
\label{Fig1}
\vspace*{-5mm}}\end{figure}

\begin{figure*}[ht]
\centering

\includegraphics[width=\linewidth]{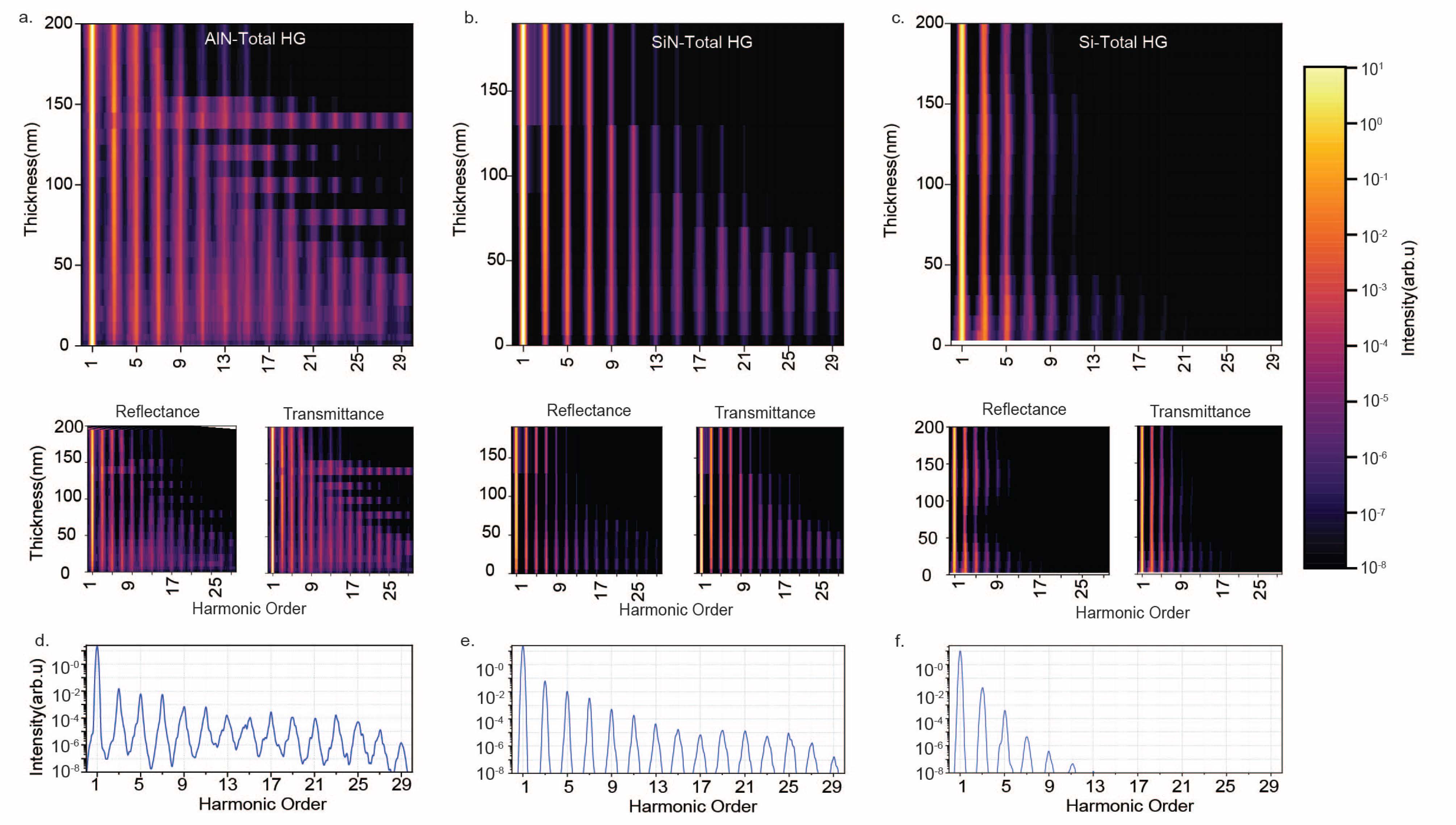}
\caption{The HHG spectra of $w$-AlN (left), $\beta$-SiN (center) and
  Si  (right). The top row of panels displays the 3D
  false-color maps visualizing the intensity of the transmitted HHG
  (on the log scale) at various thicknesses of the target (in nm). The
  middle row compares the color maps of the reflected and
  transmitted HHG. The  bottom row shows the HHG spectra at the
  target thickness of 50~nm.
\label{Fig2}
}
\end{figure*}

\begin{figure}[ht]
\centering

\includegraphics[width=\linewidth]{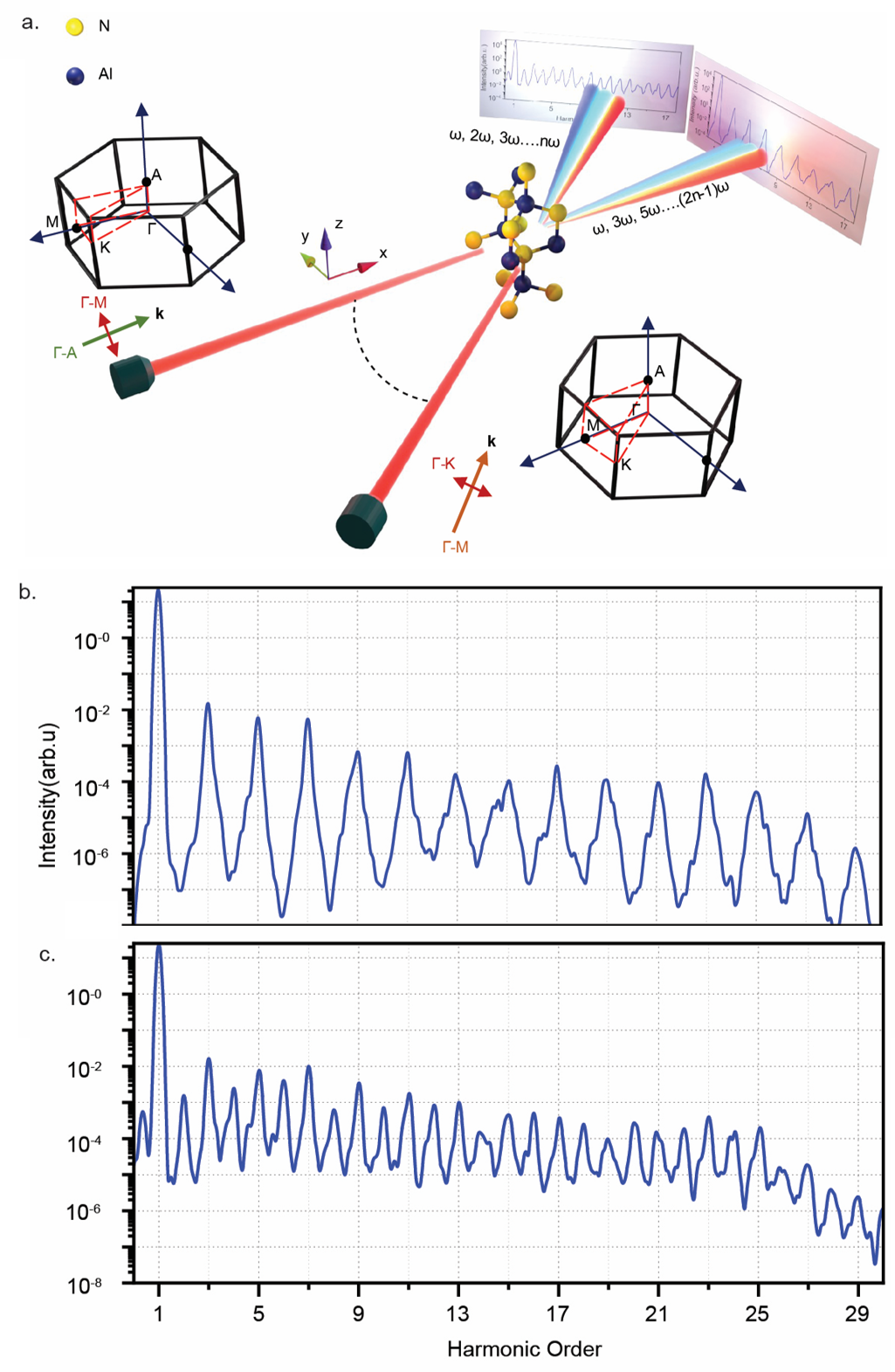}

\caption{The HHG spectra of $w$-AlN at the $\Gamma$M (left) and
  $\Gamma$K (right) alignments of the incident light polarization
  vector. The corresponding HHG spectra are displayed in the bottom
  row. 
\label{Fig3}
\vspace*{-5mm}
}
\end{figure}

High-order harmonics generation (HHG) is a nonlinear optical process
upon which a material multiplies the fundamental frequency of the
incident light. HHG arises from strong-field light matter interactions
when the electromagnetic field of light becomes non-negligible in
comparison with microscopic binding strengths of the
material. Historically, the phenomenon was associated with gas- and
plasma-phase physics, and some of the milestones were the generation
of optical harmonics up to the 11th order in plasma \cite{Burnett1977}
and 33rd harmonic generation in gas \cite{Ferray1988}.  HHG is of high
practical importance as it enables light generation in ultra-short
wavelength spectral ranges, such as extreme-ultraviolet
(XUV). HHG-based sources of XUV are currently the only devices that
fit to the table-top size. However, gas- and plasma-based systems
still find only limited applications as they are inherently bulky,
costly and require high level of professional knowledge to operate.

Over the past decade, HHG entered the realm of solid-state physics
\cite{Ghimire2019}.  All-solid-state material platforms promise
smaller and simpler systems. Physics of interactions between matter
and intense laser light in solids differs drastically from gases or
plasma, and details of the mechanisms remain a topic for debate
\cite{Ghimire2019}.  However, bulk solid-state materials (much larger
than the wavelength of light) might be able to carve only a small
niche as sources of XUV light as the bulk of all common solids absorbs
XUV. This attracted a strong interest to HHG from thin film targets
\cite{Ghimire2011, Schubert2014, Hohenleutner2015,  Luu2015,Vampa2015,Langer2017,Liu2017,You2017,Shirai2018,Vampa2018,Orenstein2019}.

The HHG studies in thin films open a unique opportunity for the solid-state platforms
due to their synergy with nanotechnology. Advances in nanofabrication
continue to drive changes in optics. Properties of materials
structured at the nanoscale can differ drastically from their bulk
counterparts opening an untapped potential for the development of new,
on-demand functionalities via elegant nano-engineering
\cite{Zubyuk2021}.  The ability to structure solids at the
subwavelength scale (nanoscale) opens new dimensions to the rich
physical landscape of interactions between solids and strong-field
light. Importantly, ultra-thin nanopatterned solids (thinner and
smaller than the wavelength of light) such as nanoresonators and
nanoresonator arrays – metasurfaces – mitigate the disadvantages of
bulk materials associated with the absorption of the XUV, attenuation
of relativistic electrons and negative impact on pulse durations. The
emerging field of HHG in nanostructured solids has several important
open questions. One of which is the choice of material
platforms. Traditionally, some of the highest-efficiency nonlinear
light-matter interactions at the nanoscale were realized in
high-refractive index semiconductors such as silicon (Si) or III-V
semiconductors including aluminium-gallium-arsenide (AlGaAs)
\cite{Koshelev2020}.  However, such materials typically have an
electronic bandgap corresponding to photon energies in the infra-red,
which hinders their ability to efficiently generate and emit XUV light.

In this work, we study theoretically suitability of wide-bandgap optical materials for generation of high order optical harmonics with
the wavelength down to the XUV spectral range. Our approach is based
on the first-principle quantum-mechanical simulations of electron
dynamics taking place during the light-matter interactions. Such
approach have seen increasing use in modelling of the HHG phenomena in
solids \cite{Otobe2016,Tancogne2017,Floss2018,Otobe2012,Yu2019,Yue22}.
In our study, we employ time-dependent density functional theory
(TDDFT), solving the Kohn-Sham (KS) equations coupled with the Maxwell
equations and solved in real time and space \cite{Yamada2021}.  Our
numerical implementation of TDDFT is based on the open-source software
SALMON \footnote{Scalable Ab initio Light-Matter simulator for Optics and Nanoscience} \cite{Noda2019,SALMON2022}. The DFT is based on the
meta-GGA exchange-correlation potential which provides for accurate
band gaps of semiconductors \cite{Tran2009}.
Using these tools, we compare HHG in thin films of the wurtzite type aluminium nitride ($w$-AlN) and hexagonal silicon nitride ($\beta$-SiN) with an analogous process from one of the most common nanophotonics materials -- silicon (Si).

\section{Simulation details}
\subsection{Materials properties}
\label{structure}
Both the target materials have the hexagonal structure as illustrated in \ref{Fig1}. The unit cell of $w$-AlN contains $2\times$Al ($2b$) and $2\times$Al ($2b$) atoms \footnote{Atomic positions are marked according to the Wyckoff crystallographic classification \cite{Bilbao2006}}. The atomic positions $2b$ are symmetric with respect to the center inversion along the $\Gamma$–M directions and
asymmetric when the coordinates are inverted along the $\Gamma$–A and $\Gamma$–K directions \cite{Kobayashi1983}.
The unit cell of $\beta$-SiN contains $6\times$Si ($6h$), $6\times$N
($6h$) and $2\times$N ($2c$) atoms \cite{Wang1996}.  These atomic
positions are fully symmetric with respect to the coordinate center
inversion.
%


Both materials are wide band gap semiconductors. The direct band gap
is 6.28~eV in $w$-AlN \cite{Nwigboji2015} and 4.5~eV in $\beta$-SiN
\cite{Belkada2000}. We will contrast these wide bandgap materials with
the diamond structure silicon $8\times$Si $(8a)$. The indirect bandgap
of Si is only 1.1~eV while the direct bandgap is 3.4~eV. This
bandgap comparison rationalizes our choice of materials for the
present study. Given the HHG cut-off at 5~eV in Si corresponding to
the band gap of 1.1~eV \cite{Vampa2018}, the same multiplication
factor of $\sim5$ would allow one to generate the harmonics in wider
band gap materials expanding well into the XUV spectral range
\footnote{One may argue that it is the direct bandgap that is relevant
  to the optical excitation and recombination in the HHG
  process. However, a low cutoff energy of Si in comparison with other
  materials displayed in Fig.~3 of \cite{Ghimire2019} implicates its
  small direct band gap that also matters.}. While a wider band gap
might be somewhat detrimental for inter-band re-collision mechanism of
HHG generation \cite{Li2023}, it would not impede the intra-band HHG
mechanism.  Moreover, a relatively narrow conduction band of about
5~eV in both materials in comparison with a 8~eV band width in Si
would promote a stronger reflection from the Brillouine zone
boundaries and would enhance a strongly non-linear HHG process.

\subsection{SALMON simulations}

Every SALMON simulation starts from the ground state calculation of
the target material in the single cell mode. This calcualtion is
served to test the electronic structure and, most importantly, the
band gap results. The corresponding values are 5.54~eV in $w$-AlN and
4.27~eV for $\beta$-SiN which is not far off from the corresponding
literature values \cite{Nwigboji2015,Belkada2000}

After the ground state properties are established, the propagation of
the incident light through the thin film is simulated using the
multi-scale Maxwell propagation technique \cite{Yabana2012}.  In this
technique, the KS equations are solved in parallel with the Maxwell
equations. While the electronic degrees of freedom are treated by the
KS equations on the microscopic scale, the Maxwell equations
describing the light propagation are solved on a much coarser
grid. This multi-scale approach allows for an efficient use of the
computational resources.  A typical calculation on $w$-AlN with the
$12\times12\times20$ partition of the unit cell and the
$4\times4\times4$ partition of the Brillouin zone would require 200
CPU-hours of the Gadi supercomputer \cite{Gadi} ranked 69 on the Top
500 list \cite{Dongarra2011}. An analogous calculation on $\beta$-SiN
with the $30\times30\times12$ partition of the unit cell would
typically require 1,600 CPU-hours. Convergence test was conducted on
$w$-AlN by refining the Brillouin zone partition to $8\times8\times8$
at the expense of a nearly an order of magnitude increase of the CPU
time. While the finer details of the HHG spectrum have changed, its
qualitative structure remained essentially the same.

Other numerical details of the present calculations are similar to the
multi-scale Maxwell simulations on Si \cite{Yamada2023}. 
The KS equations are driven by the vector-potential with the 
cos$^6$ envelope: 
\begin{equation}
    \label{eqn:1}
    A(t) = -\frac{cE_0}{\omega} \text{sin}(\omega t)
    \text{cos}^6 \left( \frac{\pi t}{T} \right), 
\ \ 
 |t| < T/2
\ .
\end{equation}
Here $\omega=1.55$~eV is the carrier frequency and $T\simeq100$~fs is
the full duration of the pulse. The amplitude of the electric field
$E_0$ corresponds to the peak intensity in the 1$\times$10$^{13}$W/cm$^2$ range.
The light propagation is described on the macroscopic scale by solving
the  wave equation:
\begin{equation}
 \left (\frac{1}{c^2} \frac{\partial^2}{\partial t^2}  -
 \frac{\partial^2}{\partial z^2} \right)  A(t)
= \frac{4\pi}{c}    J(t) \ .
\label{eq:2}
\end{equation}
The macroscopic current $J(t)$ is expressed in terms of the density of
the occupied KS orbitals and its gradient. \ref{eq:2} is solved for a
given initial pulse that is prepared in the vacuum region in front of
the thin film.  The HHG spectra are evaluated via the square of the
Fourier-transformed electric field defined as
\begin{equation}
   \tilde{E}(\omega)= \int_{0}^{ T_{\rm tot}} dt \, e^{i\omega t} E(t)
   \, f\left(\frac{t}{T_{\rm tot}}\right) \ ,
   \label{eq:3}
\end{equation}
where $T_{\rm tot}=200$~fs  is the total calculation time.

\section{Numerical results} 

The HHG spectra of the $w$-AlN and $\beta$-SiN are displayed in
\ref{Fig2} in comparison with analogous spectra of Si
\cite{Yamada2023}.  The top row of panels displays the 3D false-color
maps visualizing the intensity of the total emitted HHG (on the log
scale) at various thicknesses of the target film (in nm). These maps
display very clearly significantly higher efficiency of HHG generation in the studied
materials, especially in $w$-AlN, in comparison with Si. 
%
The middle row of panels in \ref{Fig2} details separately the HHG
portions emitted in the directions of transmission and reflection of
the excitation beam.
The bottom row of the panel shows HHG spectra at a fixed thickness of
50~nm. These panels exhibit particularly clearly a much higher HHG
output from $w$-AlN and $\beta$-SiN in comparison to Si.

In the previous section,  we noted a limited inversion symmetry of
$w$-AlN. Due to the lack of this symmetry in several directions, the
material is capable of generating both the odd and even
harmonics. These effect is illustrated in \ref{Fig3}. When the
polarization axis of the incident light is aligned with the $\Gamma$M
direction, the light-matter interaction retains the full inversion
symmetry and the corresponding HHG spectrum contains odd harmonics
only. The inversion symmetry of the light-matter interaction would
conserve the parity  and hence an odd number of the incident
light photons would be needed to convert to a single HHG photon. 
In the meantime, when the polarization axis is aligned with the
$\Gamma$M direction, this symmetry is broken and the HHG spectrum
contains both the odd and even harmonics. This effect is conceptually
similar to the observations of both even and odd HHG in wurtzite
structure ZnO\cite{WenkaiLi2022}.

\section{Conclusion}

In the present study, we have simulated high-order harmonics
generation in wide-bandgap optical materials - hexagonal $w$-AlN and
$\beta$-SiN. As a simulation tool, we used the Scalable Ab initio
Light-Matter simulator for Optics and Nanoscience (SALMON). The
light-matter interaction was simulated by simultaneous solution of the
Kohn-Sham and Maxwell equations on a vastly different scales. Such a
multi-scale approach made the simulations possible computationally in
complex materials containing a large number of atoms in their unit
cells. The present simulation demonstrates a computational and
predictive power of the SALMON package. Importantly, the present work
extends the earlier simulations on elemental solids, silicon
\cite{Yamada2023} and diamond \cite{Freeman2022}, to multi-atom
compounds.
 
The HHG spectra of $w$-AlN and $\beta$-SiN contain harmonics to a
significantly larger order when compared to a more common photonics
material -- Si.  At least 25 harmonics can be easily discerned in the
spectra of these materials in comparison with only 9 harmonics
confidently seen in silicon under similar laser pulse parameters. We
attribute this significant increase of the HHG efficiency to a
peculiarities of the band structure of the studied materials. Firstly,
a wider band gap would decrease absorption of higher order
harmonics. Secondly, a narrower conduction band would increase
non-linearity of the electron motion due to reflection from the
Brillouine zone boundaries.

An increase of the HHG efficiency makes the studied materials
promising candidates for practical applications in strong-field
nanophotonics. While other more common wide band-gap materials such as
glasses \cite{You2017} and more exotic frozen noble gases \cite{Ndabashimiye2016} allow to generate harmonics in XUV spectral range as well, $w$-AlN and $\beta$-SiN are advantageous when
synergized with nanofabrication technology and the concepts of
nanophotonics. Higher refractive index of $w$-AlN and $\beta$-SiN in
comparison to more common glasses enables intricate designs of
geometry-dependant resonances in nanostructured gratings and
metasurfaces facilitating tighter field confinements and therefore HHG
enhancements.

While the present simulation considers two- dimensional thin films, full 3D simulations of nano-structured targets are within reach with the described approach \cite{
Uemoto2019p}.
Such simulations would allow us to simulate the HHG process in grated
films and other nano-fabricated structures such as  metasurfaces bringing small, on-chip sources of XUV light a step closer.


\section{Acknowledgments} The computational resources of the NCI
Australia have been used in the present study. This work was supported
by Australian Research Council (DE210100679).

\section{Data availability} Data underlying the results presented in
this paper are not publicly available at this time but may be obtained
from the authors upon reasonable request.


\providecommand{\noopsort}[1]{}\providecommand{\singleletter}[1]{#1}%

\end{document}